# Effect of Data Sharing on Private Cache Design in Chip Multiprocessors

Leonid Yavits, Amir Morad, Ran Ginosar

**Abstract**—In multithreaded applications with high degree of data sharing, the miss rate of private cache is shown to exhibit a compulsory miss component. It manifests because at least some of the shared data originates from other cores and can only be accessed in a shared cache. The compulsory component does not change with the private cache size, causing its miss rate to diminish slower as the cache size grows. As a result, the peak performance of a Chip Multiprocessor (CMP) for workloads with high degree of data sharing is achieved with a smaller private cache, compared to workloads with no data sharing. The CMP performance can be improved by reassigning some of the constrained area or power resource from private cache to core. Alternatively, the area or power budget of a CMP can be reduced without a performance hit.

**Index Terms** — Chip Multiprocessor, Cache Hierarchy, Analytical Performance Models, Multithreaded Data Sharing

——————————— ◆ ———————————

## 1 INTRODUCTION

One of the key features of parallelizable or multi-threaded applications affecting the Chip Multiprocessor (CMP) performance is data sharing. It allows the cached data of one parallel program branch or thread to be used by other threads without the latter suffering additional off-chip DRAM access, thereby reducing the off-chip bandwidth demand. However, data sharing also requires physical transfer of large amounts of data among processing cores and throughout cache hierarchies, effectively reducing the overall CMP performance and increasing its power consumption [10][15].

It has been established that data sharing among parallel threads of a parallelizable program affect the miss rate of an on-chip cache [3][4][15]. However, prior research in this field focuses on the effect of data sharing on the last level shared cache. Moreover, majority of the existing works assume a private $L_1$ cache of constant size, access time and miss rate, and do not address the effects of data sharing on $L_1$ miss rate and size.

Our work follows the research of Krishna *et al*. [3] that studies the effect of data sharing on miss rate and size of the shared cache. Our work is different from [3] in that it focuses on the effect of data sharing on private cache.

We develop an analytical performance model that captures the relationship between the miss rate of the private cache, the cache size, and the degree of data sharing. Our modeling relies on the power law behavior of the data re-referencing rate [2], which is the foundation of the well-known $\sqrt{2}$ rule.

Our study, supported by cycle-accurate simulation, shows that while data sharing leads to miss rate reduction in the shared cache by increasing its effective size, it has an opposite effect on the private cache: with higher degree of data sharing, the impact of the private cache size on its miss rate seems to lessen. This happens because with higher degree of data sharing, the probability of the shared data to be found in the private cache of a given core decreases. One implication of this effect is a limited improvement of the overall cache performance when the private cache grows beyond a certain size. Consequently, in a constrained resource scenario, the CMP performance can be improved by shifting the resource (power, area) from cache to core.

In this work, we assume that a private cache has a single hierarchy level $L_1$. However the outcome of our research does not change if the private cache has more than one hierarchy level.

The rest of this paper is organized as follows. Section 2 describes the simulation methodology and results. Section 3 presents the analytical model of CMP performance under data sharing. Section 4 offers conclusions.

## 2 SIMULATION

In this section, we present the simulation methodology followed by simulation results.

### 2.1 Methodology

Our research platform is based on the GEM5 simulator [11]. We simulate a CMP comprising $n$ identical processing cores ($n$ is typically 12), each equipped with a private $L_1$ data cache, and a last level $L_2$ shared data cache. We use a single core CPU, equipped with a $L_1$ data cache of the same size, and a last level $L_2$ data cache of the size scaled down by $n$ as a baseline.

We simulate a number of workloads form the PARSEC 2.1 suite [5], selected by the data sharing and exchange degree (Table 1 in [5]). As we discovered during the simulation, the degree of data sharing in PARSEC 2.1 workloads is marginally sufficient to significantly affect the miss rate of a private $L_1$ cache. To better demonstrate the effect of data sharing, we also simulate a large scale FFT

————————————————
- *Leonid Yavits, E-mail: yavits@tx.technion.ac.il.*
- *Amir Morad, E-mail: amorad@tx.technion.ac.il*
- *Ran Ginosar, E-mail: ran@ee.technion.ac.il.*

*Authors are with the Department of Electrical Engineering, Technion-Israel Institute of Technology, Haifa 32000, Israel.*

workload [16].

We run each benchmark in $n$-threaded mode and create checkpoints at the start of the main work loop. Each simulation is run for 1 billion instructions or till the end of the parallel section, whichever happens sooner. For every benchmark, and every configuration we vary the $L_1$ size from 4KB through 128KB.

We analyze the effect of data sharing on $L_1$ miss rate by comparing the average $L_1$ miss rate in a $n$ core CMP vs. the $L_1$ miss rate in a single core CPU. If the data sharing is not significant enough, the miss rate behavior as a function of cache size should follow a similar pattern in both configurations. However, if there is a significant data sharing, the probability of the referenced data to be found in a local $L_1$ diminishes. Hence, the compulsory miss component unaffected by the cache size becomes more significant, so that the resulting miss rate of a private cache experiences a slower descent or even saturates with the growing cache size.

### 2.2 Results

Fig. 1 presents the average miss rate of private $L_1$ caches of the $n$ core CMP ($n$=12) as a function of their size for a number of PARSEC workloads (blackscholes, dedup and ferret) and a $2^{25}$-point FFT. For comparison, Fig. 1 also shows the miss rate of a private $L_1$ cache of a single core CPU. The PARSEC workloads have been selected based on the degree of their data sharing and exchange. Blackscholes is an embarrassingly parallel workload with virtually zero data sharing and exchange. Dedup and ferret are on the higher end of the data sharing scale of the PARSEC 2.1 workloads [5]. FFT has the highest degree of data sharing among the simulated workloads.

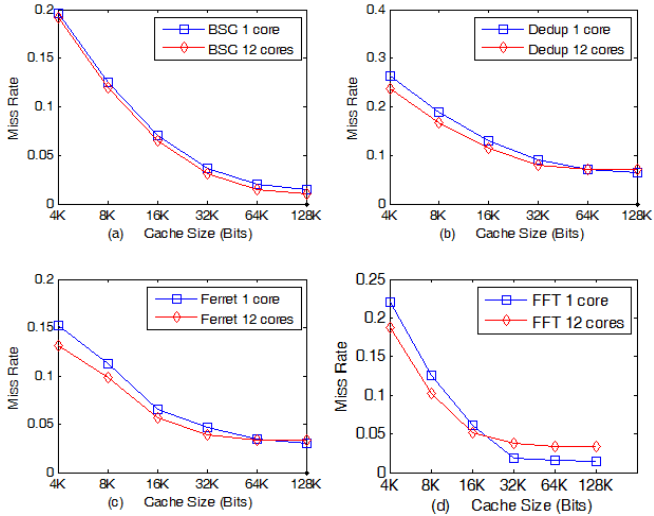

Fig. 1. $L_1$ Miss Rate vs. Cache Size: (a) Blackscholes, (b) Dedup, (c) Ferret, (d) FFT

For blackscholes, the average $L_1$ miss rate of the $n$ core CMP is very close to the $L_1$ miss rate of a single core CPU, exhibiting similar trend of diminishing with the cache size. For ferret and dedup, the average $L_1$ miss rate of the $n$ core CMP diminishes slower as $L_1$ size grows, and saturates slightly above the $L_1$ miss rate of the single core CPU. For FFT, the average $L_1$ miss rate of the $n$ core CMP exhibits much slower descent and saturates at a higher level than the $L_1$ miss rate of the single core CPU.

The reason for the difference in $L_1$ miss rate behavior is the use of the shared data. Blackscholes uses virtually no shared data, hence its cache misses are of capacity and conflict nature. Dedup and ferret both have a higher degree of data sharing, which means that some of the data required by a certain core is originated from other cores. Therefore such data can only be found in the shared $L_2$ cache. This adds a compulsory component to the $L_1$ miss rate. Such compulsory component does not depend on the cache size (the $L_1$ is to miss regardless of its size if the referenced data is sourced from elsewhere). Therefore the descent of the $L_1$ miss rate with growing size is slower. This effect is even more predominant in FFT.

## 3 ANALYTICAL MODEL FOR CMP PERFORMANCE

In this section we present an analytical model of a private cache miss rate affected by data sharing, and extend a CMP performance model to include the impact of such data sharing.

### 3.1 Private Cache Miss Rate Model

Our $L_1$ miss rate is based on the well-known $\sqrt{2}$ rule [2]:

$$m_1 = \mu/\sqrt{A_1/\alpha}; \qquad (1)$$

where $\mu$ and $\alpha$ are the miss rate and size of a baseline cache and $A_1$ is the $L_1$ cache size. Hence, we assume that the rate of data (private as well as shared) re-referencing in $L_1$ follows the power law [2].

To reflect the effect of data sharing on $L_1$ miss rate, we modify the $\sqrt{2}$ rule by adding a compulsory miss rate component $\mu_n$, as follows:

$$m_1 = \mu_n + (1-\mu_n)\mu/\sqrt{A_1/\alpha}; \qquad (2)$$

where $\mu_n$ is the compulsory miss rate component.

Fig. 2 compares the analytical models (1) and (2) to the simulation results for the FFT workload. Our model (2) seems to explain the $L_1$ miss rate behavior in the presence of shared data. The model might be limited in not accounting for all possible types of private and shared data, which may not follow the $\sqrt{2}$ rule. However we believe it provides a reasonable fit to the empiric results for a wide variety of workloads and $L_1$ cache sizes.

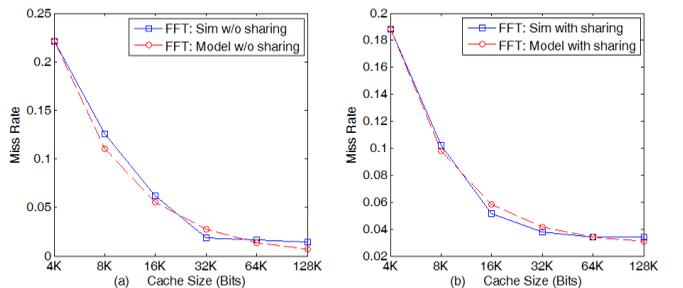

Fig. 2. $L_1$ Miss Rate Model vs. Simulation: (a) without data sharing, (b) with data sharing

The compulsory miss rate component does not depend on the $L_1$ cache size, but it may depend on the number of cores $n$: with larger $n$, the probability of a shared variable to be sourced from a different core is higher; hence $\mu_n$ is also higher.

### 3.2 Extending the CMP Performance Model

We now present the analytical model that accounts for the effect of data sharing on CMP performance.

We follow the methodology set by a variety of prior studies [1][3][6][8][9][14]. The Cycle per Instruction (CPI) of a single core reference CPU can be presented as follows [14]:

$$CPI_1 = g \cdot CPI_M + (1-g) \cdot CPI_C \quad (3)$$

where $g$ is the fraction of memory access instructions, assumed to be 0.2, $CPI_C$ is the average number of cycles per instruction for instructions that require no memory access, and $CPI_M$ is the average number of cycles per memory access.

$CPI_M$ can be presented as follows [8]:

$$CPI_M = (1-m_1)d_{L1} + m_1(1-m_2)d_{L2} + m_1 m_2 d_D \quad (4)$$

where $m_1$ (defined in (2)) and $m_2$ are the miss rates of $L_1$ and $L_2$ respectively, $d_{L1}$ and $d_{L2}$ are the access times of $L_1$ and $L_2$ respectively, and $d_D$ is the off-chip DRAM access time (including the interconnect queuing delay).

The miss rate of the $L_2$ can be written as follows [2][3]:

$$m_2 = E_n m_1 \sqrt{nA_{L1}/A_{L2}} \quad (5)$$

where $A_{L1}$ and $A_{L2}$ are the areas of $L_1$ and $L_2$ respectively, and $E_n$ is the $L_2$ data sharing impact factor [3]. Cache access times are defined as follows [8]:

$$d_{L1} = \tau \cdot \left(A_{L1}/\alpha\right)^\beta; d_{L2} = d_{NoC} + \tau \cdot \left(A_{L2}/\alpha\right)^\beta \quad (6)$$

where $\tau$ is the access time of a baseline cache and $\beta$ is a power law exponent, $\cong 0.4$ [8]; $d_{NoC}$ is a Network on Chip (NoC) delay.

$CPI_C$ can be written as follows [1]:

$$CPI_C = \frac{\chi}{\sqrt{A_{CPU}}} \quad (7)$$

where $\chi$ is the CPI of a baseline core, assumed to be 1, and $A_{CPU}$ is the area of the processing core.

Since we simulate the parallel fraction of each workload, Amdahl's parallelization factor $f$ is 1, and the performance of a $n$ core CMP can be written as follows:

$$IPC_{CMP} = \frac{n}{CPI_1} \quad (8)$$

We model under constrained area budget $A$, so that the following is always upheld:

$$n \cdot (A_{L1} + A_{CPU}) + A_{L2} = A \quad (9)$$

Two additional constraint resources that we consider in our analysis are the power budget and off-chip memory bandwidth. The power constraint can be written as follows:

$$n \cdot (P_{L1} + P_{CPU}) + P_{L2} \leq P_{max} \quad (10)$$

where $P_{L1}$, $P_{L2}$ and $P_{CPU}$ are the power consumptions of $L_1$ cache, $L_2$ cache and the processing core, respectively; $P_{max}$ is the maximum power budget available to the CMP. The cache power scales as a square root of its area, while the core power scales proportionally to its area [1]. For simplicity, we do not consider the power consumption of the NoC although it could be quite significant (almost 30% of the chip's power supply [13]).

The way to restrict the off-chip memory traffic is by limiting the rate of access to the off-chip DRAM $M_D$:

$$M_D = m_1 m_2 \leq M_{D\,max} \quad (11)$$

where $M_{D\,max}$ is the maximum bandwidth capacity.

Fig. 3 presents the $IPC_{CMP}$ as a function of the area budget, with and without the data sharing (the latter one is calculated identically to the former one, while setting $\mu_n=0$ in (2)). The workload with no data sharing achieves a higher performance than the workload with a high degree of data sharing [10][15], because it requires much less time-consuming data traffic throughout the cache hierarchies. In both constrained budget scenarios (off-chip bandwidth and power), the CMP performance $IPC_{CMP}$ exhibits retrograde behavior, as the area overcommitment leads to longer delays and higher power consumption [10][12][15].

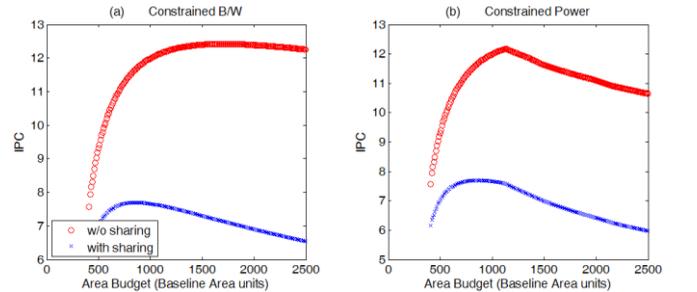

Fig. 3. $IPC_{CMP}$ vs. Area budget under: (a) constrained off-chip B/W, (b) constrained power.

One outcome of the performance analysis is that the optimal $L_1$ area point is different for workloads with high degree of data sharing vs. those with no data sharing. This is demonstrated in Fig. 4, which presents the $IPC_{CMP}$ as a function of $L_1$ area $A_{L1}$. The function is discontinuous since for every $A_{L1}$ value, there are number of possible $A_{L2}$ and $A_{CPU}$ values, leading to different performance figures. The maximum performance of a workload with no data sharing is reached at considerably higher $A_{L1}$ than that of a workload with a high degree of data sharing. The difference in optimal $A_{L1}$ reaches 80% for the constrained off-chip bandwidth scenario, and 40% for the constrained power budget scenario, respectively. This result confirms the notion that a significant portion of $L_1$ area or power budget can be reallocated elsewhere in the presence of shared data.

In scenarios modeled in this work, the area budget is fixed (9), hence $L_1$ area has to be reallocated to either $L_2$ or the processing core. It has been shown that data sharing also leads to a smaller $L_2$ [3], hence core is the ultimate target. However, in power-constrained environment, increasing the shared $L_2$ is the least power-consuming option, hence the growing area budget is being assigned to $L_2$, with a retrograde effect on the CMP performance. When the off-chip bandwidth is the dominant constraint, the excess $L_1$ area is divided between core (thus improving core performance) and $L_2$ (decreasing the $L_2$ miss rate and thus reducing the rate of off-chip DRAM access $M_D$). However, increasing the area budget beyond the optimal point leads to performance degradation in both scenarios.

Practically, the area (and consequently the power consumption) of a CMP should be limited by not increasing the $L_1$ (as well as $L_2$) area beyond the point of optimal performance.

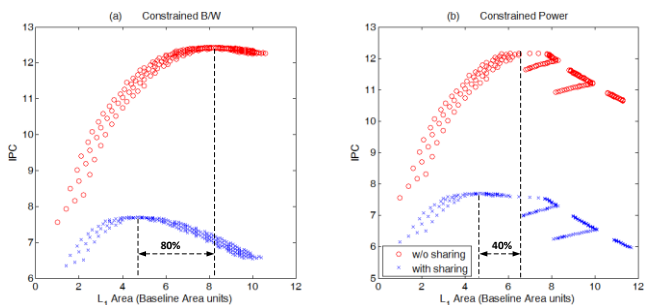

Fig. 4. $IPC_{CMP}$ vs. $L_1$ area under: (a) constrained off-chip B/W, (b) constrained power.

## 4 CONCLUSIONS

Data sharing in parallelizable or multi-threaded workloads may become a source of CMP performance degradation and excessive power consumption – because of the need to physically transfer large amounts of data among processing cores and throughout cache hierarchies. However if the effect of data sharing on cache miss rate is carefully quantified, such performance degradation can be limited.

In this work we characterize the effect of data sharing in parallelizable workloads on miss rate of private cache and on overall performance of a CMP. We show that data sharing leads to a compulsory miss component in the miss rate of a private cache. Such compulsory miss component does not decline with the cache size, thus putting a limitation on the private cache size's impact on the overall CMP performance. We extend the CMP analytical performance model to incorporate the effects of data sharing on miss rate of private cache. We show how the optimal size of private cache decreases when the data sharing is introduced, while optimizing the CMP performance under constrained bandwidth and power resource.

We find that high degree of data sharing can significantly impact the optimal design points for CMP. In particular, we find that data sharing allows a significantly larger fraction of the area to be reallocated from private cache to core. Alternatively, the area and power budget of a CMP can be reduced without a performance hit.

## ACKNOWLEDGMENT

This research was partially funded by Intel Collaborative Research Institute for Computational Intelligence and Hasso-Plattner-Institut.